# Comparison of Graphene Formation on C-face and Si-face SiC {0001} Surfaces

Luxmi, N. Srivastava, Guowei He, and R. M. Feenstra
Department of Physics, Carnegie Mellon University, Pittsburgh, PA 15213
P. J. Fisher
IBM T. J. Watson Research Center, Yorktown Heights, NY 10598

**Abstract**
The morphology of graphene formed on the ($000\bar{1}$) surface (the C-face) and the (0001) surface (the Si-face) of SiC, by annealing in ultra-high vacuum or in an argon environment, is studied by atomic force microscopy (AFM) and low-energy electron microscopy (LEEM). The graphene forms due to preferential sublimation of Si from the surface. In vacuum, this sublimation occurs much more rapidly for the C-face than the Si-face, so that 150°C lower annealing temperatures are required for the C-face to obtain films of comparable thickness. The evolution of the morphology as a function of graphene thickness is examined, revealing significant differences between the C-face and the Si-face. For annealing near 1320°C, graphene films of about 2 monolayers (ML) thickness are formed on the Si-face, but 16 ML is found for the C-face. In both cases, step bunches are formed on the surface and the films grow continuously (carpet-like) over the step bunches. For the Si-face in particular, layer-by-layer growth of the graphene is observed in areas between the step bunches. At 1170°C, for the C-face, a more 3-dimensional type of growth is found. The average thickness is then about 4 ML, but with a wide variation in local thickness (2 – 7 ML) over the surface. The spatial arrangement of constant-thickness domains are found to be correlated with step bunches on the surface, which form in a more restricted manner than at 1320°C. It is argued that these domains are somewhat disconnected, so that no strong driving force for planarization of the film exists. In a 1-atm argon environment, permitting higher growth temperatures, the graphene morphology for the Si-face is found to become more layer-by-layer-like even for graphene thickness as low as 1 ML. However, for the C-face the morphology becomes much worse, with the surface displaying markedly inhomogeneous nucleation of the graphene. It is demonstrated that these surfaces are unintentionally oxidized, which accounts for the inhomogeneous growth.

## I. INTRODUCTION

Epitaxial graphene on SiC{0001} has been intensively studied over the past five years as a potential means of producing *large area* graphene for electronic applications.[1] There are two inequivalent faces of SiC{0001} – the (0001) face, also known as the Si-face, and the ($000\bar{1}$) face or C-face. On both surfaces graphene can be formed by annealing the SiC in vacuum, causing preferential sublimation of the Si atoms thereby leaving behind excess C atoms which self-assemble into the graphene. Structural studies have revealed many differences between the graphene formation on the Si-face and the C-face: Graphene forms easily on the C-face *i.e* thicker films are formed on the C-face than on the Si-face for a given annealing time and temperature.[2,3] A complex 6√3×6√3-R30° interface layer (denoted 6√3 for short) exists between graphene and the SiC for the Si-face,[4,5,6,7,8] whereas the interface structure is quite different (but not so well understood) for the C-face.[9,10,11] This 6√3 layer appears to act as a template layer for graphene on the Si-face, ensuring well ordered graphene on that surface, but no such template layer forms for the C-face.[11] The unit cell of graphene films on the Si-face are rotated by 30° with respect to the SiC substrate, as seen by low-energy electron diffraction (LEED), but the C-face LEED shows streaking due to rotational disorder in the graphene layers.[12,13] This rotational disorder has been shown to produce electronic properties of *multi-layer* graphene films (on the C-face) that resemble those of *single-monolayer* graphene, even for films consisting of many monolayers.[14] Recently, preparation of graphene in an argon[15,16] (or disilane[17]) environment has been shown to improve the structural quality of the film, particularly for the Si-face.

In this work we show that the *morphology* of graphene formed on the C-face by annealing in vacuum is very much different than that found for a similar graphene thickness on the Si-face. Atomic force microscopy (AFM) and low-energy electron microscopy (LEEM) are used to demonstrate that graphene on the C-face forms with a somewhat 3-dimensional (3D) morphology, displaying step bunches separating domains of multi-layer graphene, with quite different thicknesses between neighboring domains. In contrast, for the Si-face the graphene forms in a more layer-by-layer manner. We argue that this difference arises in some way from the lower temperature used for the graphene growth on the C-face, inhibiting coarsening of neighboring graphene domains. In particular, we provide evidence that on the C-face the adjacent graphene domains, for formation temperatures <1250°C, are not well connected across step bunches, so that there is not a strong driving force for adjacent domains to planarize.

In an effort to increase the growth temperature for the C-face, while maintaining a fixed growth rate, we study the graphene formation in a 1-atm argon environment. In that case, we achieve essentially perfect layer-by-layer growth of the graphene on the Si-face,[18] in agreement with prior reports.[15,16] In contrast, for the C-face we find in the initial stage of formation that the graphene forms 3D islands having thickness ≥5 monolayers (ML) before coalescence occurs, also in agreement with recent works.[19,20] The C-face surface in argon thus appears to be resistant to uniform graphitization, suggesting that it is relatively low energy (*i.e.* stable). This conclusion is surprising considering that the C-face graphitizes relatively easily in vacuum, as mentioned above, indicative of a high energy (unstable) nature. However, we demonstrate that the surface when annealed in the argon environment is unintentionally oxidized, and it is this oxidation that gives rise to the island growth.

## II. EXPERIMENT

Our experiments are performed on nominally on-axis, *n*-type 6H-SiC or semi-insulating 4H-SiC wafers purchased from Cree Corp, with no apparent differences between results for the two types of wafers. The wafers are normally 2 or 3 inch in diameter, mechanically polished on both sides and epi-ready on either the (0001) surface or the ($000\bar{1}$) surface. These wafers are cut into 1×1 cm$^2$ samples and the samples are chemically cleaned in acetone and methanol before putting them into our custom built preparation chamber which uses a graphite strip heater for heating the samples.[21] Samples are first etched in a 10 lpm flow of pure hydrogen for 3 min at a temperature of 1600°C. This H-etching removes the polishing scratches which arise during the mechanical polishing of the wafers, resulting in an ordered step-terrace arrangement on the surface which is suitable for graphene formation.[22] Before annealing, hydrogen is pumped away from the chamber and we wait until a desired pressure of 10$^{-8}$ Torr is reached. The samples are then either annealed in this vacuum for 10 to 40 min at temperatures ranging from 1100-1400°C, or under 1 atm of flowing argon (99.999% purity) for 15 min at ≈1600°C. Temperature is measured using a disappearing filament pyrometer, with calibration done by using a graphite cover over the sample and measuring its temperature.[21]

Following graphitization the samples are transferred to an Elmitec LEEM III system for LEEM and LEED measurements. Samples are outgassed at 700°C prior to study. The sample and the electron gun are kept at a potential of -20 kV and LEEM images are acquired with electrons having energy, set by varying the bias on the sample, in the range of 0-10 eV. The intensity of the reflected electrons from different regions of the sample is plotted as a function of the beam energy. These LEEM reflectivity curve shows oscillations, with the number of graphene monolayers (ML = 38.0 carbon atoms/nm$^2$) being given by the number of local minima in the curve.[23] From sequences of images acquired at energies varying by 0.1 eV, color-coded maps of the graphene thickness are generated using the method described in Ref. [18].

The vacuum system containing the LEEM is also equipped with a 5 kV electron gun and VG Scientific Clam 100 hemispherical analyzer used for Auger electron spectroscopy (AES). For routine determination of graphene thickness by AES we use the ratio of the 272 eV KLL C line to the 1619 eV KLL Si line. This ratio is analyzed with a model involving the escape depths of the electrons,[24] with the overall magnitude of the ratio being calibrated to graphene thicknesses determined by LEEM. (The model includes for the interface one ML of carbon, *i.e.* the 6√3 layer, for the Si-face, but no such layer for the C-face).

LEED patterns were acquired not only in the LEEM system but also using a separate ultra-high-vacuum system containing a VG Scientific LEED apparatus. Sample were transferred to that system through air (*i.e.* an *ex situ* measurement). In a few experiments, this LEED apparatus was connected to the annealing chamber so that LEED measurement could be performed *in situ*. The surface morphology of the graphene films was studied in air by AFM, using a Digital Instruments Nanoscope III in tapping mode.

### III. RESULTS
### A. Graphene formation in vacuum

Figure 1 shows the plot of graphene thickness as a function of annealing temperature for graphene grown on the C-face and Si-face of SiC. In agreement with prior reports,[2,3] we find that graphene starts to form at a significantly lower temperature on the C-face compared to the Si-face, about 1100°C for the former and 1250°C for the latter. At a given temperature, thicker graphene film forms on the C-face than the Si-face. It can be seen that about 9 ML are formed on the C-face at about 1250°C while only one monolayer is formed on the Si-face. The data for the Si-face is in agreement with our previous report for that surface, with an uncertainty of ±50°C in the temperatures which can be attributed to difficulties associated with measuring the temperature of a transparent sample accurately with optical pyrometry.[21]

Graphene films have a fixed rotational orientation with respect to the SiC substrate on the Si-face as seen in their LEED patterns which display a hexagonal arrangement of six clear, distinct spots rotated by 30° relative to the SiC spots.[12,13,25] However on the C-face, there is a rotational disorder in the graphene layers giving rise to streaking in the diffraction pattern,[14] as illustrated in Fig. 2. We still observe six discrete spots in the pattern, seen weakly at the maximal angles indicated in the figure, with these spots located at 30° relative to the SiC spots (the latter are not seen in these patterns, but clearly apparent in data for thinner graphene films). However, additional spots (streaks) are also seen located at angles of 30° ± φ relative to the six discrete spots. Angles of φ ranging from 6 to 13° have been observed, although most typically we find φ ≈ 7°. Overall our patterns are quite similar to those reported by Hass et al.,[14] although their angle φ was only 2.2°. We interpret this difference as arising from the various possible nearly-commensurate arrangements of graphene monolayers atop each other.[26]

LEEM and AFM studies have been performed to monitor evolution in morphology for graphene on C-face SiC. In this respect it is important to note that, as discussed in Ref. [27], graphene formation on the C-face is found to depend significantly on the surface properties of the starting SiC wafer. If, after H-etching, the surface displays a well-ordered array of parallel, straight steps edges (arising from unintentional miscut of the surface), as in Fig. 1(b) of Ref. [28], then high quality graphene will form. Alternatively, if the surface displays numerous spiral step arrays (associated with screw dislocations intersecting the surface), as in Fig. 6 of Ref. [28], or some other type of poorly-ordered step array, then we find that significant amounts of what is believed to be nano-crystalline graphite (NCG) is formed. In the worst case, this NCG can cover the entire surface, although generally it appears only near step edges over some fraction of the surface. We believe that this formation of the NCG is related to the inhomogeneous nucleation of graphene discussed by Camara et al.,[19,29] who observe both intrinsic and extrinsic graphene formation, with the latter arising from dislocations intersecting the surface. Their extrinsic graphene forms in an ordered manner, whereas we find disordered NCG, but the growth temperatures employed by Camara *et al.* are considerably higher than ours and we believe that that could account for this difference. In any event, all of the LEEM results in this section were obtained from samples that displayed well-ordered steps after H-etching and hence high quality (intrinsic) graphene films.

Figure 3(a) shows an AFM image of 6H-SiC($000\bar{1}$) graphitized at 1100°C for 20 min. The surface has preserved the uniform step-terrace arrangement as seen after H-etching, with terraces showing small domains of varying gray-contrast. These steps can also be discerned in the LEEM image shown in Fig. 3(b). This image shows the reflected electron intensity, at an electron energy of 3.3 eV. As described by Hibino *et al.*,[23] areas of the graphene with different thickness interact differently with the incident electrons, thus producing varying contrast for the graphene films as a function of energy. Plots of the reflected intensity as a function of energy are shown in Fig. 3(c), for the specific locations A-G indicated in Fig. 3(b). Secondary electrons produce the large reflectivity below about 1.5 eV. The number of minima in the curves A-D above that energy corresponds to the local thickness in ML of the graphene film. Thus, for curve B, acquired from the distinctly white contrast region of Fig. 3(b), there are 2 ML of graphene. This type of contrast extends over roughly half of Fig. 3(b), and the remainder having a darker, but somewhat mottled, contrast. Reflectivity curves from those types of areas reveal a combination of 1 ML and 3 ML graphene coverage, as illustrated by the curves A and C. Finally, a few small regions of this surface reveal 4 ML graphene thickness, as shown by the curve D. Regions marked E-G show no well-defined oscillations, as seen in the reflectivity plot of Fig. 3(c). These regions might contain very small domains (<50 nm) of varying thicknesses. LEEM images acquired at different beam energies are analyzed pixel by pixel to generate a color-coded map of local graphene thickness, as shown in Fig. 3(d). It can be seen that the sample is covered with domains of different graphene thicknesses, mainly 2 ML.

It is important to realize that, as the graphene forms, the surface of the sample will *recede* since Si atoms are leaving[10] (assuming limited interdiffusion of the Si and C atoms, as demonstrated below). The carbon content in a single graphene monolayer (38.0 atoms/nm$^2$) is very close to that in three SiC bilayers (36.5 atoms/nm$^2$). The latter constitutes 0.75 nm of height in its SiC form, whereas the graphene monolayers are spaced by about 0.34 nm from each other and have similar spacing to the SiC (for the C-face) or the 6√3 layer (for the Si-face).[30] Thus, for each additional ML of graphene, the top surface must recede by about 0.4 nm. The data of Fig. 3(d) showing a mixture of 1, 2, 3 and 4 ML graphene thickness is consistent with the 0.4-1.2 nm variation in surface height across the terraces on this surface as seen in Fig. 3(a) [an exceptionally low region is marked by the arrow in Fig. 3(a) and it is likely associated with the 4 ML graphene thickness].

With further annealing, the morphology of the surface changes. Figure 3(e) shows an AFM image of a surface graphitized at 1150°C for 20 min. Its surface morphology is quite different than that of Fig. 3(a). We now see that the step edges are somewhat irregular, with flat regions of the surface now forming irregularly-shaped μm-sized regions separated from their neighboring terraces by step bunches. Traces of the original step edges from the H-etched surface are still visible in the Fig. 3(e); these traces are small deposits, likely carbon in the form of NCG, that form at the step edges during the initial graphitization and these persist even during subsequent graphitization. To emphasize the difference in surface morphology between Fig. 3(a) and 3(e), we show in Fig. 4 AFM results from two other samples prepared in a similar manner to each of those. Again, the image of Fig. 4(a), prepared at relatively low temperature, shows a uniform step array with slightly varying contrast along each terrace [due to varying thicknesses of graphene, as in Fig. 3(a)], whereas the image of Fig. 4(b), prepared at higher temperature, shows the irregularly-shaped surface terraces, similar to Fig. 3(e). The terraces are separated by step bunches, ≈3 nm high. It is clear that the change in temperature from ≈1100 to ≈1170°C produces significant motion of the steps on the surface, although this motion is still quite limited compared to what occurs at 1320°C as discussed below.

Examining now the morphology of the graphene film for the surface prepared at 1150°C, Figs. 3(f) – 3(h), we find domains with 1-3 μm lateral extent and having a wide range of graphene thicknesses, from 2 ML to 7 ML. We note that the reflectivity curves provide a faithful measure of the graphene thickness over this entire range. In Fig. 5 we compare the measured location of the minimum of the reflectivity curves together with a simple prediction based on a tight-binding model for a one-dimensional chain,[31] using the same parameters as in Ref. [23] but with central energy of 3.2 eV above the vacuum level (0.2 eV higher than used in Ref. [23], but that work discusses the Si-face whereas our data is for the C-face). We can see that there is a reasonably good match between theory and experiment. Below 1.5 eV the reflectivity curves show an intense shoulder due to secondary electrons, which affects the location of minima near that energy. Even accounting for that, the match between experiment and theory is still somewhat better for the upper half of the curves (above 3.2 eV) than for the lower half. However, as pointed out by Hibino *et al.*, this discrepancy is likely due to the fact that the true bandwidths differ above and below this central energy (an effect that is not present in the tight-binding model, but *is* seen in first-principles computations).[23]

The situation we find for the C-face, with graphene domains having a very wide range of thicknesses is much different from that found for the Si-face, discussed in the following Section, for which only a 1 ML range in graphene thickness is found over most of the surface. This large range in graphene thicknesses for the C-face, essentially a 3D growth phenomenon, is expected to have significant deleterious effects on electrical behavior of the films and it is therefore of interest to identify the source of this three-dimensional morphology. One possibility is that the limited step motion on the C-face surface could, in some way, lead to the limited lateral extent of the graphene domains. In particular, there could possibly be some correlation between the location of the *surface terraces*, Fig. 3(e), and the *graphene domains*, Fig. 3(h), and thus one might be influencing the other. To further investigate this we have performed AFM and LEEM imaging over identical surface areas.

Figure 6(a) shows a LEEM image acquired with 8.6 eV electron energy and Fig. 6(b) shows an AFM image of the *same* surface region. A large defect located off of the right-hand side of the images, together with smaller defect indicated by the arrows and trenches indicated by crosses, permit precise alignment of the images. Analyzing the graphene thickness, a color-coded map is shown in Fig. 6(c). The areas colored gray and white have graphene thicknesses of 7 and 8 ML, respectively. These areas were outlined by visual inspection, and dashed lines of those outlines are superimposed on the AFM image of Fig. 6(d). We see that these areas of thick graphene correspond reasonably quite well to the areas of lower (darker) surface height. We thus conclude that there is indeed limited interdiffusion of C and/or Si within the SiC during the graphitization, so that regions from which Si has left then convert locally into graphene, and the surface height recedes accordingly. The data of Fig. 6 shows ≈2 nm lower (darker) surface regions where the graphene is locally several ML thicker than its surroundings, which is consistent with our expected 0.4 nm drop in surface height per graphene ML.

We find that many of the edges of the graphene domains in Fig. 6(c), particularly those that correspond to ≥2 ML of thickness change, are located at the position of step bunches as seen in Fig. 6(d). This correlation is illustrated in Fig. 6(e), where we show cross-sectional cuts of the AFM topography along the lines indicated in Fig. 6(d). Lines at the same location are indicated in Fig. 6(c), and from that data we obtain the local graphene thickness. The solid lines of Fig. 6(e) show the surface topography from AFM, and the dashed lines show the depth of the graphene below the surface (in many cases on the lower terrace there are small deposits of NCG on the surface, as discussed above, and we ignore that NCG in placing the dashed lines for the graphene/SiC interface). For cut C-C, across a ≈3.5-nm-high step bunch, the graphene thickness is 5 ML on the upper terrace and 7 ML on the lower. For B-B, across a ≈4.0 nm bunch, the graphene thickness is 3 ML on the upper terrace (directly adjoining the step bunch) and 5 ML on the lower. For A-A, across a ≈3.2 nm bunch, the graphene thickness is 6 ML on the upper terrace and 4 ML on the lower (although along a cut slightly above A-A the thickness is 4 ML on both sides of the step bunch). In all cases, the depth of the graphene on the upper terrace is less than the height of the step bunch, and this relationship also holds true for most other step bunches found on the surface.

The correlation found between the locations of the step bunches and the boundaries between graphene constant-thickness domains suggests that, indeed, the formation of the step bunches and the graphene domains are somehow coupled. Apparently the limited step motion on the surface at these temperatures (< 1200°C) also produces some limitation in the extent of the graphene domains. It appears from the data that adjacent graphene domains do not planarize (*i.e.* establish coincident upper surfaces and/or lower graphene/SiC interfaces), and this inability to planarize seems to be related to the existence of the step bunches at the boundaries between domains.

The "domains" that we have discussed above refer to areas of constant graphene thickness. These are not necessarily the same as a grain size (*i.e.* crystallographic domain size) in the graphene, and indeed, prior measurements indicate that the grain size is much smaller than our domain size, typically ≤100 nm.[32] To check this value for our own samples, we have performed selected-area LEED. We employed a sample similar to that of Fig. 3, with 4.3 ML average graphene thickness and having constant-thickness domains with lateral extent of 3-4 μm. We acquired LEED patterns using an illumination aperture of the LEEM that corresponded to an area on the sample of 2 μm in diameter. The patterns resembled those of Fig. 2. Examining the patterns at scores of locations over the surface, spaced by about 5 μm from each other, we do *not* find any observable variation from location to location in the pattern. This result implies that either the crystallographic orientation is unchanged over the surface, or that the grain size is much smaller than the 2 μm sampling size. The former is impossible since the streaks themselves observed in the LEED patterns demonstrate multiple orientations of the grains, so we conclude, consistent with prior works, that the grain size is much less than 2 μm. As previously discussed,[27] we do observe in our samples at low graphene coverage 200-nm-sized areas that display varying contrast in AFM, and these areas likely are different crystallographic domains in the graphene.

Thus far we have discussed graphene on C-face films with average thickness less than about 4 ML, formed by annealing at temperature <1200°C. Higher temperature annealing produces thicker films, and an AFM image of one such film, formed at 1320°C. is shown in Fig. 7. Step bunches, 3-6 nm high, are clearly visible in the topographic cut taken through the image. A striking difference between this image and those of our ≈4 ML graphene films [Figs. 3(e), 4(b), or 6(b)] is that Fig. 7 displays very prominent ridges (white lines) in the image, arising from strain relaxation of the film. These features are characteristic of films that extend continuously (carpet-like) over the step bunches so that elastic strain relaxation of the film produces the raised ridges.[1,33] The continuity of the ridges over step edges in Fig. 7 provides evidence of the continuous growth of the graphene over the edges [indeed, transmission-electron microscopy (TEM) of such films directly reveals this type of carpet-like growth of step edges[34]]. In contrast, our graphene films prepared at ≈1170°C do not display any such strain-induced ridges.

We take the absence of the ridges in the 1170°C-prepared films to indicate that the strain in these films is relieved in some other manner, *e.g.* at boundaries between the constant-thickness domains. Coupled with our observation that many of those domain boundaries occur at step bunches, we conclude that the films may well be somewhat discontinuous at the bunches. This discontinuity does not necessarily mean that *no* graphene is formed at the step bunches, but rather, just that the carpet-like growth of the graphene over the boundaries that occurs for the higher-temperature C- and Si-face films does not appear to proceed so well for these 1170°C-prepared C-face films.

**B. Graphene formation in argon**

To achieve a narrower distribution of the thickness-domains on the C-face, while maintaining a relatively thin film, it seems clear that higher formation temperatures in an argon (or disilane) environment is needed. This method has been demonstrated to reduce the sublimation rate of the Si, thus permitting an increased temperature without developing a thick graphene film. The method works very well on the Si-face, where annealing in 1 atm of Ar at 1600°C is found to produce large domains of single thickness (*e.g.* 1 ML) graphene.[15,16,18] We have attempted in eight experimental runs to form thin graphene on the C-face under 1 atm of argon, using nominally similar preparation conditions (≈1600°C for 15 min) each time. About half of those attempts resulted in nearly no graphene at all (as detected by AES), and the other half produced very thick (>15 ML) graphene films. However, in two cases for samples that displayed no graphene over most of their surface, there were a few isolated 0.1-mm-sized areas that were graphitized. These areas are easily visible under an optical microscope.

AFM and LEEM studies near the edge of one such area are shown in Fig. 8. In the AFM image, Fig. 8(a), there are many ridges (white lines at various angles) extending over the surface on the right and left-sides of the images. These features are well known to be characteristic of the presence of graphene on the surface, and they arise from the mismatch in thermal expansion coefficients between the graphene and the SiC as discussed in the previous Section. However, near the center of the image (to the right of the step bunch) no such ridges are seen, thus suggesting that no graphene is present there. This inhomogeneous coverage of the graphene is consistent with the AES measurements just mentioned, and also consistent with the LEEM results described below.

Figure 8(b) shows a LEEM image acquired at 5.2 eV, and reflectivity curves from the associated sequence of images are shown in Fig. 8(c). Over most of the surface we find the same sort of reflectivity curve as presented in Fig. 3, with the number of minima providing a measure of the graphene thickness. Curves C-G correspond to 1 − 5 monolayers, respectively. Curve C actually has an additional shallow minimum, marked by the dashed line at 6.8 eV, and this same feature is weakly seen in curve D. But, other than that, the other minima in all the curves match up very well with the results already presented in Fig. 3 [the curves in Fig. 8(c) are shifted upwards by about 0.5 eV, but this can be the result simply of a different alignment of the electron beam in the LEEM]. A color-coded map of the graphene thicknesses in shown in Fig. 8(d), revealing an average graphene thickness (over the area covered by graphene) of 3.0 ML.

On the left-hand side of the LEEM image of Fig. 8(b) is seen a black region, with reflectivity given by curve A. The reflectivity is seen to be nearly featureless over the range 3 − 10 eV, without the characteristic oscillations of the graphene. It should be noted in this regard that, in addition to the oscillations in the range 2 − 7 eV, the reflectivity from graphene also increases over the energy range 8 − 10 eV because of additional band structure effects.[35] This increase at higher energies is also not seen for curve A. The same reflectivity as in curve A was found over the vast majority of the surface. Thus, we can be certain that the surface, at location A in Fig. 8(b) and over the vast majority of the sample, is not covered with any graphene at all.

Figure 9 provides additional LEEM results from the same sample, with Figs. 9(a) and 9(b) showing data acquired near the edge of an island, and Figs. 9(c) and 9(d) showing data acquired near the center of an island. These results are consistent with those of Fig. 8, revealing the average thickness of the graphene islands of about 4.1 ML near the island edge and increasing to 4.6 ML near the island center. The anomalous minimum near 6.8 eV is also seen in the reflectivity curves for 1 ML and 2 ML thickness near the edge of the island, Fig. 9(b), although not at the center of the island, Fig. 9(d).

Returning for a moment to Fig. 8(c), the reflectivity curve B has a shape never before seen by us nor reported by others. This reflectivity curve exists only over the small area colored black in Fig. 8(d) near location B, although we have found identical curves on other sample areas. The origin of this new reflectivity as well as the extra minima seen in the 1 and 2 ML curves is not known at present, although we note that the latter resemble the additional minima produced in reflectivity curves obtained on Si-face graphene when it is intercalated by H.[36] Oxygen has been used in a similar manner on the Si-face.[37] As described below, our samples annealed in argon turn out to be unintentionally oxidized, with a silicate layer ($Si_2O_3$) on their surface. Perhaps this silicate layer is affecting the reflectivity curves and producing the new features we observe, although the details of this effect are not understood at present. In any case the main conclusion from the data of Figs. 8 and 9 is clear: This surface, prepared at high temperatures under 1 atm of argon, is covered only in a few areas by graphene, and there the graphene is many ML thick. Elsewhere on the surface no graphene is present. Thus, we find islanding of the graphene, similar to that reported recently by both Camara *et al.*[19] and Tedesco *et al.*[20]

LEED obtained from areas of the Ar-annealed samples that do not have any graphene display clear SiC 1×1 spots together with faint √3×√3-R30º spots (the latter vary in intensity over the surface). This same pattern is found for the measurement performed *ex situ* or *in situ*. In Fig. 10(b) we display one of these patterns and compare it to a 3×3 LEED pattern formed by annealing a C-face sample in vacuum, Fig. 10(a). The surfaces prepared in vacuum or argon are clearly very different. We have measured LEED intensity *vs.* energy spectra for the √3×√3-R30º pattern, as shown in Fig. 10(c). The results agree very well with the known spectra for a silicate ($Si_2O_3$) layer on SiC($000\bar{1}$),[38] with residual oxygen present during the Ar annealing apparently oxidizing the surface.

However, it should be noted in this regard that, in vacuum, the silicate layer is unstable at temperature above about 1200°C, at least for the Si-face[39]. This fact raises the possibility that the oxidation observed on our argon-annealed sample might have occurred while the sample was cooling down to room temperature, or during evacuation of the Ar gas. To investigate this we have taken a C-face 3×3 surface formed by annealing in vacuum, exposed it for 10 min at various temperatures to a 1-atm Ar environment, and measured the resulting LEED pattern. For room temperature annealing we find that the LEED pattern becomes noticeably dimmer but that the 3×3 spots are still faintly visible; no trace of any √3×√3-R30º spots are seen. But, after annealing in the Ar to > 1000°C, the √3×√3-R30º spots appear. This pattern grows markedly in intensity as the temperature is increased to 1200°C, and then it maintains an approximately constant intensity as the temperature is increased to 1550°C. For annealing at 1640°C we find that the surface is graphitized over most of its area, although a few regions of intense √3×√3-R30º remain. Thus, we find that the silicate is stable, in the Ar environment, for temperature up to ≈1600°C.

For the C-face in vacuum we found that it graphitizes easier than the Si-face, indicating a higher surface energy of the C-face. Now, in argon, we find that the C-face surface is more resistant to graphitization than the Si-face, indicative of a lower surface energy for the C-face. The presence of the oxide layer on the C-face surface accounts for this difference in the surface energies between the vacuum and argon environments, thus providing an explanation for the difficulty in graphitizing the C-face in argon. Apparently the C-face is more sensitive to this type of contamination than is the Si-face.

**IV. DISCUSSION**

Our previous results for graphene formation on the Si-face of SiC, as a function of graphene thickness, have been presented in Refs. [18] and [21]. Those results are in agreement with prior reports of other workers.[1-7,9-14] Before graphene forms on the Si-face, the well known 6√3 layer forms on the surface after annealing in vacuum at temperatures of ≈1200°C.[40] Further heating to about 1250°C produces graphene thickness of 1 – 2 ML (Fig. 1). At this stage the surface morphology is rather disordered, with many small pits on the surface (as a result of the 6√3 formation[5]) and considerable small-scale motion (≈1 μm) of surface steps. The graphene morphology is also rather disordered, with small domains (< 1 μm lateral extent) of 1 and 2 ML graphene thickness present on the surface. Further annealing up to ≈1350°C causes much longer range motion of the surface steps, with step bunches forming separated by ≥10μm. At the same time, the graphene domains coarsen and grow to areas with lateral extent of many microns. Essentially, the areas between step bunches are covered with graphene ranging in thickness over only 1 ML. Hence, for thicknesses > 2 ML, the graphene is found to form in a layer-by-layer manner (i.e. away from the step bunches).

For our *in vacuo* results, Fig. 3, it appears that the initial graphene formation on the C-face is not so different than on the Si-face. In both cases the initial lateral extent of constant-thickness (1 or 2 ML) domains of the graphene is ≲100 nm. With subsequent annealing the C-face graphene morphology coarsens, forming large areas (several μm) with 2 ML coverage, as well as small areas with 3 ML coverage. This process continues, with the several-μm-sized domains of graphene becoming thicker. However, the lateral extent of the domains does not further increase for the C-face, even for thickness up to 8 ML [Fig. 3(h)], and the range of thicknesses on the C-face is about 5 ML whereas it is limited to a single ML (away from step bunches) for the Si-face.

If we compare the Si-face and C-face graphene morphologies for a fixed film thickness, then they are very different, as just described. But if we instead compare them at fixed temperatures, the differences become understandable. At ≈1320°C, the films thickness on the C-face is much greater than for the Si-face (16 *vs.* 2 ML), but both films display the characteristic ridges associated with strain relaxation and both surfaces display comparable amounts of step bunching. The reason for the thicker film on the C-face is, we believe, simply because the ($000\bar{1}$) surface and ($000\bar{1}$)/graphene interface have higher energies (*i.e.* are more unstable), respectively, than the (0001) surface and (0001)/graphene interface. Additionally, more defects in the C-face films such as the discontinuities discussed in Section III(A) and/or rotational domain boundaries could lead to easier Si diffusion through the graphene, which would also favor thicker growth.[41,42] Turning to the C-face graphene prepared at ≈1170°C, and comparing that to the higher-temperature C-face graphene, it seems clear that the 1170°C-prepared material has lower structural quality, due to kinetic limitations from the reduced growth temperature. Thus, the different morphologies between the Si- and C-faces found for films of the same thickness simply arises from the lower graphene formation temperatures used in the latter case, which inhibits coarsening between adjacent domains. We believe that the fundamental growth mode is 2D, *i.e.* with the graphene wetting the SiC surface, for both the Si-face and the C-face.

When graphene is formed on the Si-face under 1 atm of argon, the tendency to grow in a layer-by-layer manner become even more pronounced.[15,16,18] In that case, it is quite easy to produce a single ML extending over 10's or 100's of μm on the surface, with longer annealing (or higher temperatures) presumably leading to a second ML, etc. The contrast between that situation and what we find for the C-face is stark. For the C-face under argon, as shown in Figs. 7 and 8, we find island formation of the graphene with high growth rate once an island is nucleated. However we have demonstrated that this C-face surface is oxidized, and this oxidation apparently inhibits the graphene formation. The islands observed at temperatures near 1600°C arise either from areas of the surface from which the oxide decomposes, or they are related in some way to the extrinsic growth of graphene (from dislocations) as discussed by Camara *et al.*[19] Similarly, the thick films we observe at higher temperatures arise from either (or both) of these sources. In any case, we are unable to use the Ar annealing method to vary the growth temperature in order to controllably study few-ML films of graphene formation on the C-face.

Recently, Tedeso *et al.* reported on island formation of graphene on the C-face under an Ar environment.[20] Our results are in agreement with theirs, at least when the Ar is present. But, for annealing in vacuum they report that they also observe island formation of the graphene. This differs from our results of Section III(A), in which we find continuous graphene films that wet the surface, *i.e.* a 2D growth mode, albeit one with somewhat nonuniform growth due to kinetic limitations. We speculate that the island growth observed in by Tedesco *et al.* for their *in vacuo* studies may arise from some unintentional contamination of their surface due to their background pressure of only $10^{-5}$ Torr, since we have found that the C-face is relatively sensitive to such contamination. For the radio frequency induction furnace (pressure unspecified) used by Camara *et al.*[19,29] some unintentional species in the background gas might also be present, which could possibly account for the higher growth temperatures needed by those workers compared to us.

**V. SUMMARY**

In summary we have studied the formation of graphene in UHV on the SiC ($000\bar{1}$) surface (the C-face) using AFM and LEEM. By comparison of the results with our prior measurements for the (0001) surface (the Si-face) we are able to understand certain aspects of the formation kinetics. For *in vacuo* preparation and coverage of < 2 ML we find that the morphology of the graphene is similar on both surfaces, comprising small areas 100's of nm in extent of given thickness. We refer to these areas of fixed thickness as *domains* of the graphene. Further formation of the graphene produces coarsening of the domains, but the process is very different on the C-face compared to the Si-face. For the C-face the domains grow laterally up to an extent of only several microns. Importantly, for average thickness of 4 ML, the variation in thickness over the surface is quite large, with thinnest and thickest regions differing by 5 ML. The graphene thus forms a 3-dimensional type of morphology. In contrast, for the Si-face at coverage > 2 ML, the graphene domains coarsen considerably forming areas with extent of many microns or more. The variation in thickness over the surface (away from step bunches) is limited to 1 ML, so that the graphene is seen to form in a layer-by-layer manner.

For a fixed graphene thickness, the morphology of steps on the surface are quite different between the C-face and the Si-face, with the ≈150°C higher graphitization temperature for the Si-face leading to much greater step motion. On the C-face, with formation temperatures of about 1170°C yielding average graphene film thicknesses of ≈4 ML, we observe characteristic terraces in the surface morphology with lateral extent of 1-3 μm. The size of the graphene constant-thickness domains seen in LEEM is similar, and we suggest a connection between the two. Correlated AFM and LEEM imaging reveals that many of the boundaries between constant thickness domains (particularly for relatively thick domains) occur at the location of the step bunches, from which we propose that the graphene films are somewhat discontinuous at the step bunches. This discontinuity would then tend to inhibit the formation of a more uniform thickness distribution in the C-face graphene film, since the disconnected areas have no driving force for forming a flat, continuous graphene/SiC interface across domains. We thus argue that the 3D morphology found for C-face graphene is a consequence of kinetic limitations due to its relatively low growth temperature.

We have studied the graphene formation under a 1 atm argon environment. For the Si-face in argon, the layer-by-layer growth mode is more firmly established, with the growth of a single monolayer of graphene over 10's or 100's of μm being relatively easy to achieve. But for the C-face in argon, only 3D formation of islands is found in the initial stage of graphene formation, with these islands growing relatively thick (≥5 ML) before complete graphene coverage is achieved. This stark difference between the two surfaces (in contrast to the situation in vacuum for which the C-face graphitizes *easier* than the Si-face) implies some difference in the surface structure of the C-face under a vacuum or argon environment. LEED observations of the surface structure under the two conditions do indeed yield different results, with the annealing under argon producing unintentional oxidation of the surface.

**ACKNOWLEDGEMENTS**

This work was supported by the National Science Foundation, grant DMR-0856240. Discussions with Gong Gu, J. E. Northrup, M. G. Spencer, and U. Starke are gratefully acknowledged.

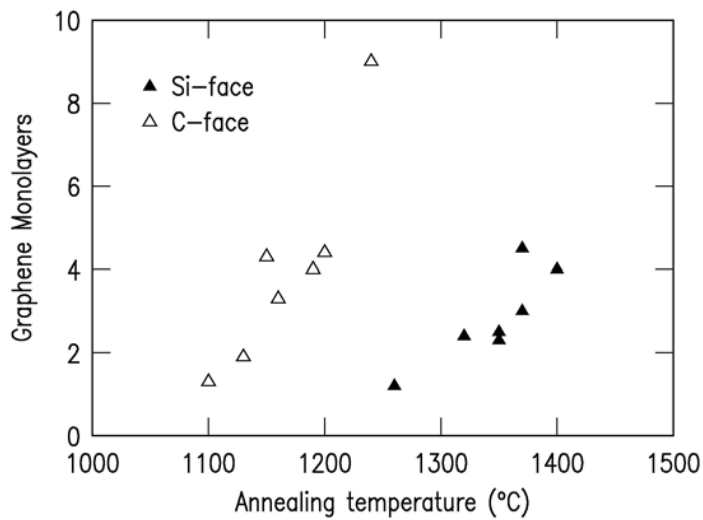

FIG. 1. Graphene thickness as a function of annealing temperature for 6H-SiC {0001} surfaces, showing results for C-face (anneal time 20 min) and Si-face (anneal time 40 min).

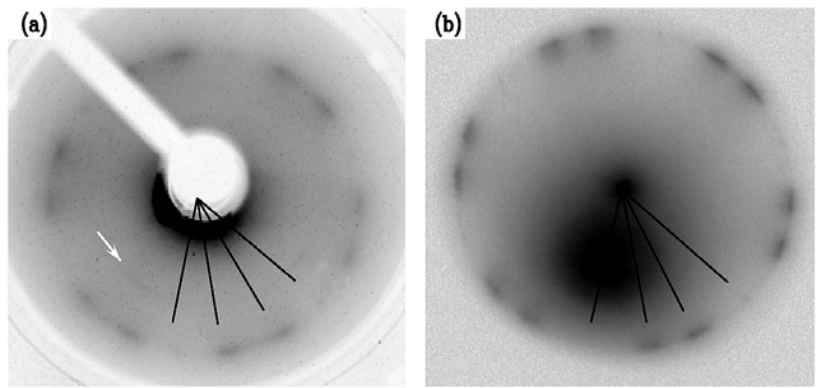

FIG. 2. LEED pattern for graphitized SiC($000\bar{1}$) surfaces, obtained at the energies of (a) 133 eV (using a VG Scientic LEED apparatus) and (b) 44 eV (using an Elmitec LEEM III). Samples were prepared by annealing for 20 min, (a) 4H-SiC($000\bar{1}$) surface at 1370°C and (b) 6H-SiC($000\bar{1}$) surface at 1235°C. The faint streak indicated by the arrow in (a) is an artifact (optical reflection) of the video acquisition system. The black lines indicate diffraction features over a 60° range of angles, with the outer peaks of this range being oriented at 30° relative to primary SiC (1,0) spots.

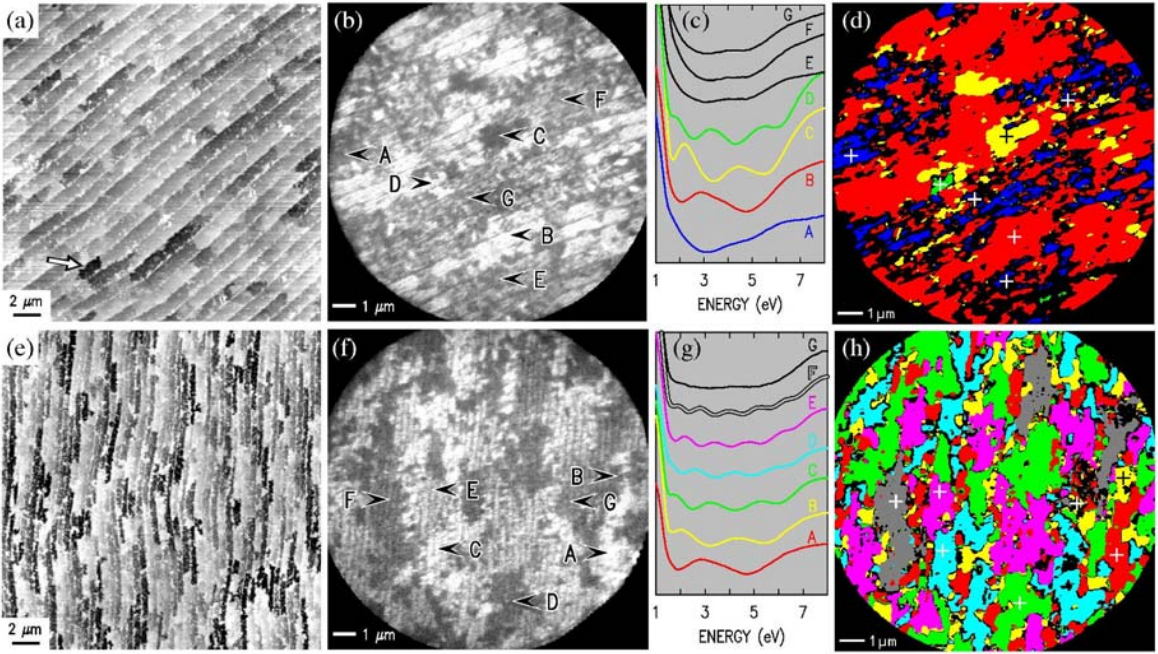

FIG. 3. (Color online) Results from graphitized 6H-SiC($000\bar{1}$) surfaces prepared by heating in vacuum under conditions of (a)-(d) 1100°C for 20 min, yielding an average thickness of 2.0 ML of graphene, and (e)-(h) 1150°C for 20 min, yielding 3.9 ML of graphene. (a) and (e) 20×20 um² AFM images, displayed with gray scale ranges of 3 and 4 nm, respectively. (b) and (f) LEEM images at an electron beam energy of 3.3 eV with 15 μm field-of-view. (c) and (g) Intensity of the reflected electrons from different regions marked in (b) or (f) as a function of electron beam energy (curves are shifted vertically, for ease of viewing). (d) and (h) Color-coded maps of local graphene thickness, deduced from analysis of the intensity vs. energy at each pixel; blue, red, yellow, green, cyan, magenta, and gray correspond to 1 − 7 ML of graphene, respectively. Small white or black crosses mark the locations of the intensity vs. energy curves. Regions with no discernable oscillations are colored black.

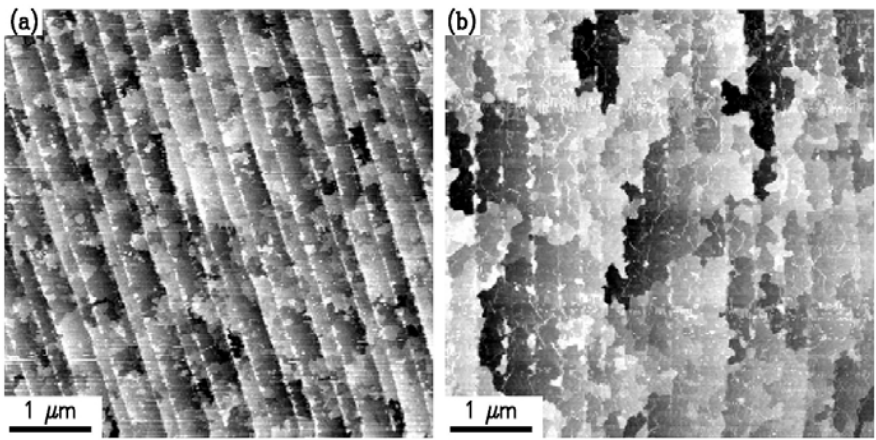

FIG. 4. AFM images of graphene formed on 6H-SiC($000\bar{1}$) surfaces by annealing for 20 min at (a) 1120°C and (b) 1190°C. Resulting graphene thicknesses are (a) 1.2 ML and (b) 4.0 ML. Gray scale ranges for the images are (a) 2 nm and (b) 4 nm.

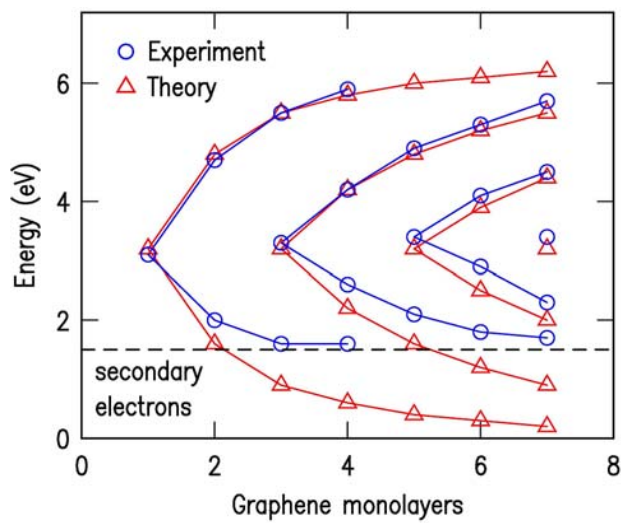

FIG 5. (Color online) Location of local minima in the reflectivity curves from Fig. 3(g), compared with theoretical expectations based on a tight-binding model (see text).

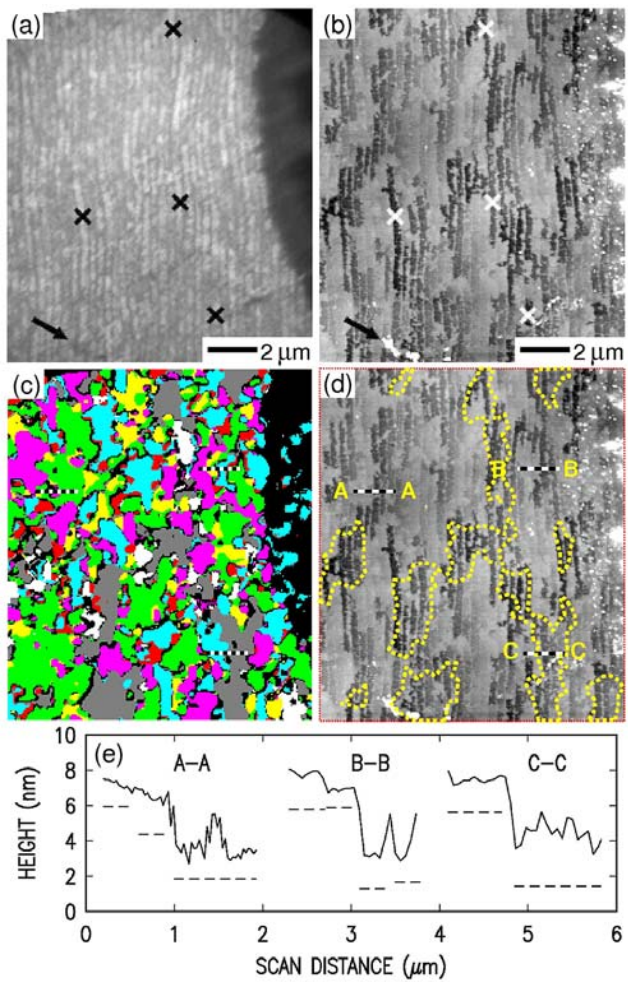

FIG 6. (Color online) AFM and LEEM images acquired from identical surface areas of a graphitized 6H-SiC($000\bar{1}$) surface [same sample as in Figs. 3(e) – 3(h)]. (a) LEEM image acquired at 8.5 eV. (b) AFM image, with relative surface height represented by a gray scale ranging from 0 nm (black) to 10.5 nm (white). Arrows in (a) and (b) indicate a surface defect, and crosses indicate trenches in the surface morphology. (c) Color-coded map of local graphene thickness, presented in same manner as Fig. 3(h) and with white areas indicating 8 ML of graphene. (d) Same AFM image as (b), but now with areas of gray and white from (c) indicated by yellow dashed lines. (e) Solid lines show cross-sectional cuts along the black/white dashed lines indicated in (d). Black/white dashed lines at identical locations are also shown in (c), revealing the local thickness of the graphene, as indicated by dashed lines in (e).

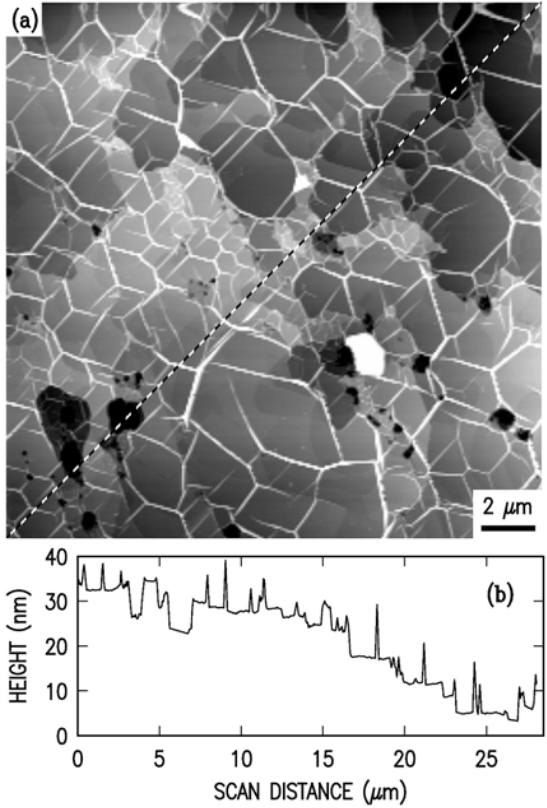

FIG 7. (a) AFM image of graphene film prepared on 6H-SiC($000\bar{1}$) by annealing at 1320°C for 20 min, resulting in an average film thickness of 16 ML. (b) Topographic cut, along the dashed line in (a).

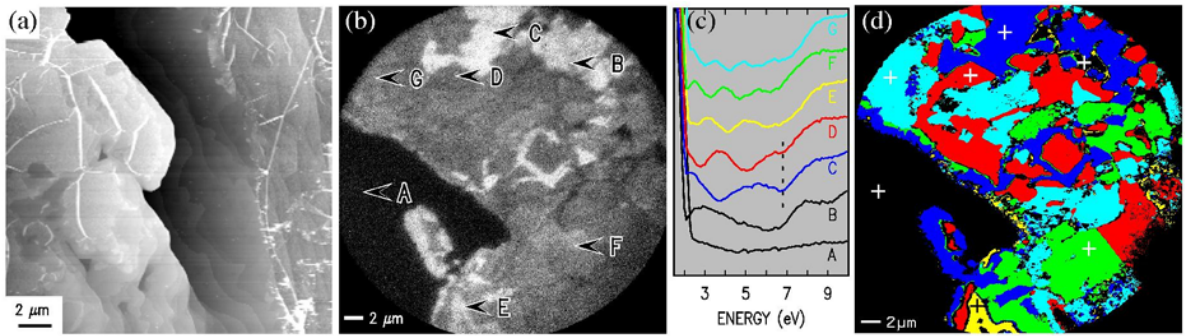

FIG 8. (Color online) Results of graphene prepared on SiC($000\bar{1}$), by annealing at 1600°C for 15 min in 1 atm of argon yielding an average thickness of 3.0 ML of graphene (for this image, including only the areas where graphene covers the surface). (a) AFM image with gray scale range of 16 nm, (b) LEEM image at beam energy of 5.2 eV and with 25 μm field-of-view. (c) Intensity of the reflected electrons from different regions marked in (b) as a function of electron beam energy, and (d) Color-coded map of local graphene thickness, presented in same manner as for Figs. 3(d) and 3(h).

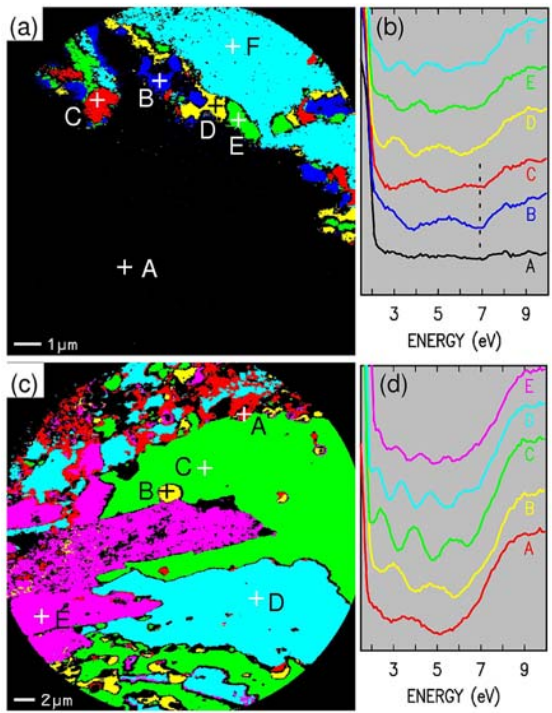

FIG 9. (Color online) LEEM results from the same sample and using the same manner of display as Fig. 8, showing results near the edge of a graphene island [(a) and (b)] with average graphene thickness of 4.1 ML in this portion of the island, and near the center of an island [(c) and (d)] with average graphene thickness 4.9 ML. (a) and (c) Color-coded maps of local graphene thickness; (b) and (d) reflectivity curves acquired from the locations indicated in (a) and (b), respectively.

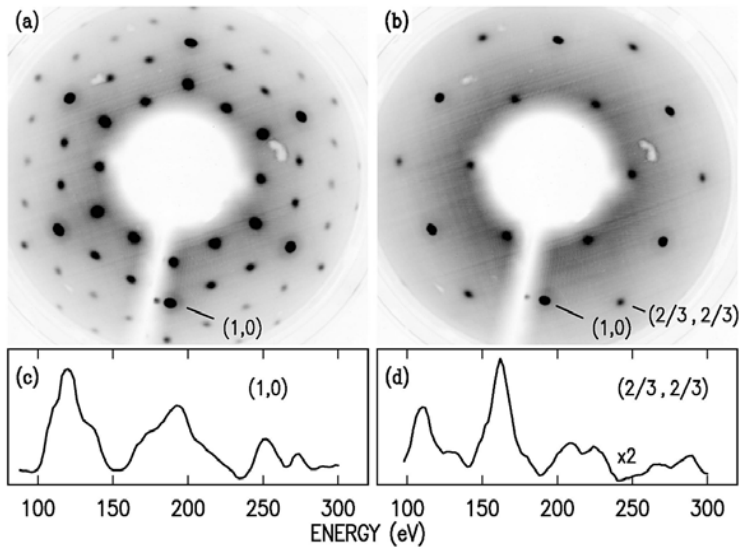

FIG 10. LEED data acquired from SiC($000\bar{1}$) surfaces: (a) 3×3 pattern acquired at 100 eV from a sample prepared by annealing at 1000°C in vacuum, with the primary SiC (1,0) spot indicated; (b) √3×√3-R30° pattern acquired at 100 eV from a sample prepared by annealing in 1-atm argon at 1400°C, with the (1,0) and (2/3, 2/3) spots indicated; (c) and (d) Intensity *vs.* energy characteristics for the two spots marked in (b).